# Cell-sized confinements alter molecular diffusion in concentrated polymer solutions due to length-dependent wetting of polymers

Yuki Kanakubo[1], Chiho Watanabe[2], Johtaro Yamamoto[3], Naoya Yanagisawa[1], Arash Nikoubashman[4, 5], Miho Yanagisawa[1,6,7*]

[1] Komaba Institute for Science, Graduate School of Arts and Sciences, The University of Tokyo, Komaba 3-8-1, Meguro, Tokyo 153-8902, Japan
[2] School of Integrated Arts and Sciences, Graduate School of Integrated Sciences for Life, Hiroshima University, Kagamiyama 1-7-1, Higashi-Hiroshima 739-8521, Japan
[3] Biomedical Research Institute, National Institute of Advanced Industrial Science and Technology (AIST), Central 6, Higashi 1-1-1, Tsukuba, Ibaraki 305-8568, Japan
[4] Institute of Physics, Johannes Gutenberg University Mainz, Staudingerweg 7, 55128 Mainz, Germany
[5] Department of Mechanical Engineering, Keio University, Hiyoshi 3-14-1, Kohoku, Yokohama 223-8522, Japan
[6] Graduate School of Science, The University of Tokyo, Hongo 7-3-1, Bunkyo, Tokyo 113-0033, Japan
[7] Center for Complex Systems Biology, Universal Biology Institute, The University of Tokyo, Komaba 3-8-1, Meguro, Tokyo 153-8902, Japan

*Corresponding Author:
myanagisawa@g.ecc.u-tokyo.ac.jp (M. Y.)

**ABSTRACT**

Living cells are characterized by the micrometric confinement of various macromolecules at high concentrations. Using droplets containing binary polymer blends as artificial cells, we previously showed that cell-sized confinement causes phase separation of the binary polymer solutions because of the length-dependent wetting of the polymers. Here we demonstrate that the wetting-induced heterogeneity of polymers also emerges in single-component polymer solutions. The resulting heterogeneity leads to a slower transport of small molecules at the center of cell-sized droplets than that in bulk solutions. This heterogeneous distribution is observed when longer polymers with lower wettability are localized at the droplet center. Molecular simulations support this wetting-induced heterogeneous distribution by polymer length. Our results suggest that cell-sized confinement functions as a structural regulator for polydisperse polymer solutions that specifically manipulates the diffusion of molecules, particularly those with sizes close to the correlation length of the polymer chains.

**TOC graphic**

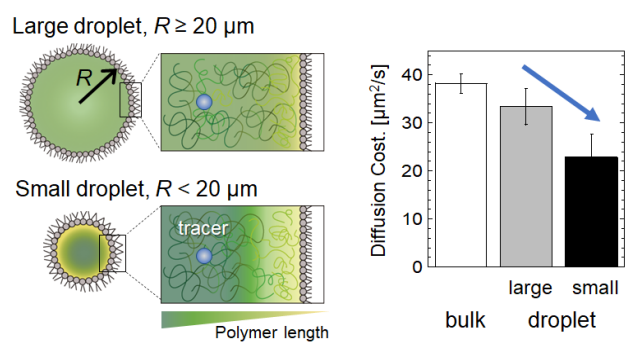

The cytoplasm of living cells typically contains numerous (macro)molecular components at high concentrations.[1-2] These components cover a wide size spectrum, ranging from a few nanometers in single proteins to several micrometers in cytoskeletal filaments. The positioning of these components within the cell is strongly tied to their biological functions, e.g., the long-lived cytoskeleton formed by intermediate filaments beneath the cell membrane exhibits strain stiffening in response to an external load, thereby protecting cells from harmful deformations.[3] The identity and concentration of the (macro)molecular components can be analyzed from the fluorescence labeling of target molecules [4] and measurements of local physical properties, such as the refractive index (RI),[5] polarization,[6] and absorption efficiency of light.[7-8]

Recently, the dynamic formation of biomolecular condensates through liquid–liquid phase separation (LLPS) has attracted attention as a mechanism to regulate the dynamic positioning of molecules. For example, in an in vitro system, Monterroso et al. demonstrated that LLPS can modulate the assembly of the protein, FtsZ, which plays a key role in bacterial cell division.[9] From a polymer physics perspective, the formation and positioning of biomolecular condensates depend on many factors, including molecular properties (e.g., the sequences of RNA nucleotides and protein amino acids); molecular concentration; and solution conditions (e.g., temperature, pH, and salinity), which have been extensively studied through experiments and simulations.[10-11] An important but frequently overlooked property is molecular length. However, it is challenging to characterize the length distribution of molecules inside and outside the biomolecular condensates owing to difficulties in multi-fluorescent labeling by length. Furthermore, physical quantities, such as RI, are less sensitive to the molecular length than to the concentration (see a case study on dextran solutions[12]).

Recently, we demonstrated that such heterogeneous distributions by polymer length have become particularly prominent for binary polymer solutions confined in cell-sized water-in-oil (W/O) droplets covered with a lipid membrane.[13-14] Experiments and simulations have revealed that shorter polymers have a higher membrane wettability, which, in turn, enhances the localization of short/long polymers at the surface/center of the droplets. This results in LLPS at concentrations where bulk systems would still be in the fully mixed one-phase regime.[14-15] Based on those observations, we hypothesized that even in the one-phase regime, spatial heterogeneity can emerge in single-component solutions of polydisperse polymers confined in small droplets owing to the length-dependent membrane wetting of polymers.[16-17]

To verify this hypothesis, we study the concentrated solutions of dextran polymers confined in small droplets using experiments and simulations in the present study. We determine the spatial distribution of dextran chains near the droplet surface for various dispersity in length and analyze the diffusion of small molecules near the center of the droplets. This situation resembles the macromolecular crowding in cell nuclei, where long DNA, and chromatin macromolecules surround small molecules, such as proteins.[18] We expect that the small molecules passing through the overlapped polymer chains change their diffusion according to the local distribution of the polymer chains.

As illustrated in Figure 1, we analyzed the molecular diffusion inside W/O droplets covered with a phosphatidylcholine (PC) lipid membrane, where 400 mg/mL dextran with a molecular weight of $M_w$ = 150k (Dex150k) was confined. The estimated mesh size, $\xi$ (or correlation length), of the dextran solution is in the range of 1.9–2.4 nm (sections S1 and S2 in the Supporting Information (SI)). It is known that the transport of particles with radius $r$ through polymer chains strongly depends on the size ratio, $2r/\xi$ (Figure 1b)[19-23]. For $2r \ll \xi$, which can be achieved for small particles or low polymer concentrations, the particle diffusivities are expected to be similar to those in a pure solvent. However, for $2r \gg \xi$, the particle diffusivities should scale as $(2r/\xi)^{-2}$, according to scaling theory[24]. In the intermediate regime, $2r \approx \xi$, the particles and polymer segments frequently collide before they can fully relax, resulting in strong coupling with sub-diffusive behavior at intermediate times. To cover a broad $2r/\xi$ regime, we chose two diffusive molecules: green fluorescent protein (GFP) with $2r$ = 5.6 nm [25] and 5-carboxytetramethylrhodamine (TAMRA) with $2r$ = 1.7 nm.[26]

First, we measured the diffusion of the TAMRA molecules ($2r/\xi$ = 0.7–0.9) by using fluorescence correlation spectroscopy (FCS). Figure 1c shows the typical autocorrelation functions (ACFs) inside the droplets and in the bulk, revealing that the TAMRA diffusion inside a small droplet with $R$ = 13 μm slows down compared with that in a large droplet with $R$ = 58 μm and in the bulk. To evaluate the observed $R$-dependent diffusion, we fitted the fractional Brownian motion (fBM) equation to the ACFs measured inside the droplets with different $R$ values (section S1 in the SI). The values of the obtained anomalous exponent, $\alpha$ are plotted against $R$ in Figure 1d; they are almost the same as the bulk value of $\alpha_{bulk}$ = 0.94 ± 0.02 (average value (Ave.) ± standard deviation (S.D.), sample number N = 25), regardless of $R$. This slightly sub-diffusive behavior is in very good agreement with that in previous simulation results of colloid–polymer mixtures, where a value of $\alpha_{bulk}$ = 0.98 ± 0.01 was found for $2r/\xi$ = 0.7.[19] It means that the diffusion mode of TAMRA in the dextran solution is almost Brownian at any droplet size examined. We also derived the long-time diffusion coefficient, $D$, from the same data assuming normal diffusion ($\alpha$ = 1) and normalized it by the bulk value, $D_{bulk}$, of 38.2 ± 2.1 μm$^2$/sec (Ave. ± S. D., N = 25) (Figure 1e). Contrary to the $R$-independent $\alpha$, $D/D_{bulk}$ decreases as $R$ decreases below ~20 μm, as previously reported.[27] This slow translational diffusion of TAMRA inside small droplets was not observed for droplets without dextran (the inset of Figure 1e), which was confirmed with our previous reports on TAMRA and Rhodamine 6G (R6G) [27], R6G in droplets containing liner polymer polyethylene glycol (PEG), and GFP in droplets containing bovine serum albumin.[25] Therefore, the observed $R$-dependent slow diffusion is a phenomenon that occurs inside small droplets crowded with polymers, regardless of the types of macromolecular crowder and tracer molecules.

Next, we measured the diffusion of GFP ($2r/\xi$ = 2.3–2.9) inside the droplets. The GFP molecules exhibit a much stronger sub-diffusive behavior both in the bulk ($\alpha_{bulk}$ = 0.82 ± 0.04 (Ave. ± S.D., N = 14)) and in the droplets with any $R$ (Figure 1f), which agrees with previous simulations where a value of $\alpha_{bulk}$ = 0.79 ± 0.01 was reported for $2r/\xi$ = 2.7.[19] Focusing on the small droplets with $R$ < 20 μm, there is a slight decrease in both $\alpha$ and $D_\alpha$ with decreasing $R$ (Figures 1f, 1g), but the $R$-dependence of $D_\alpha$ is significantly weaker than that of TAMRA (Figure 1e). These results mean that GFP molecules are much strongly trapped in the dextran chains compared with the TAMRA molecules, which is consistent with the increase in $2r/\xi$. In our previous report, we confirmed that crowder polymers also exhibit $R$-dependent slow diffusion inside small droplets containing concentrated polymer solutions.[25] Therefore, the strong coupling between GFP and polymer chains may have caused the distribution of the distance traveled by the GFP to become non-Gaussian with a long tail, such that $\alpha$ and $D$ fall with a decrease in $R$, where the polymer relaxation is also slowed.[28-29]

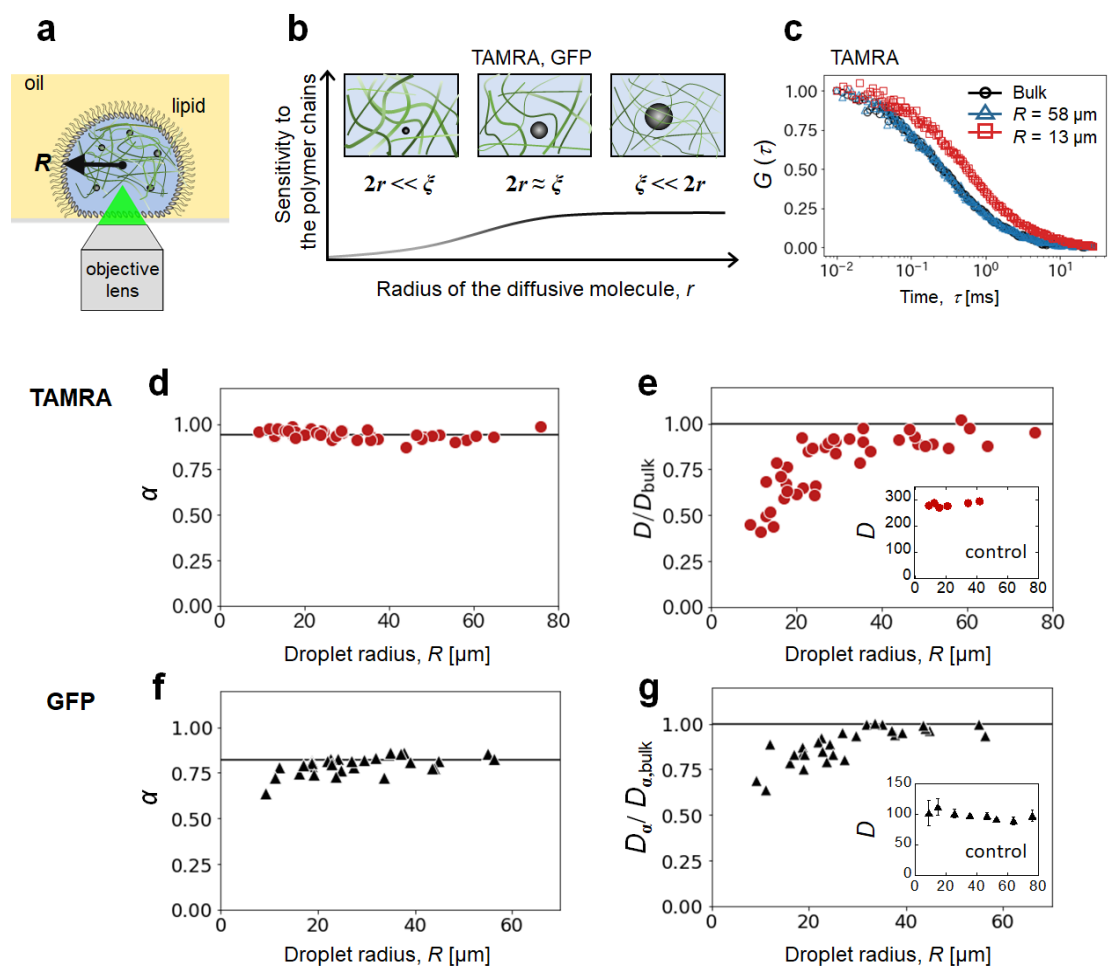

**Figure 1**. (a) Schematic of the FCS experiment for dextran PC droplets. (b) Illustration of the sensitivity of diffusive molecules to the polymer chains against the size ratio between the polymer mesh size, $\xi$, and diameter of the tracer molecule, $2r$. (c) ACFs of TAMRA diffusion inside droplets containing 400 mg/mL Dex150k (squares and triangles) and in the corresponding bulk (black circles). The values are normalized at $10^{-2}$ ms. The droplet radius, $R$, is 58 μm (blue triangles) and 13 μm (red squares). The bulk ACF (black circles) is practically identical to the ACF of the large droplet (blue triangles). (d–g) $R$-dependent diffusion inside droplets containing 400 mg/mL Dex150k for (d–e) TAMRA ($2r/\xi \approx 0.7$–$0.9$) and (f–g) GFP ($2r/\xi \approx 2.3$–$2.9$). (d, f) Anomalous exponent $\alpha$ and (e, g) normalized diffusion coefficient $D/D_{bulk}$ is plotted against $R$. Their bulk values are (d) $\alpha_{bulk} = 0.94 \pm 0.02$ (ave. ± S.D., N = 25); (e) $D_{bulk} = 38.2 \pm 2.1$ μm²/s (Ave. ± S.D., N = 25); (f) $\alpha_{bulk} = 0.82 \pm 0.04$ (Ave. ± S.D., N = 14); and (g) $D_{\alpha, bulk} = 17.6 \pm 1.6$ μm²/s (Ave. ± S.D., N = 14) for $\alpha_{bulk} = 0.82$. Insets of (e, g) are $D$ values for $\alpha = 1$ in droplets containing a pure buffer (control). Inset of (e) is reprinted with permission from ref. [27]. Copyright 2021 American Chemical Society. Inset of (g) is reprinted with permission from ref. [25]. Copyright 2020 American Chemical Society.

Unlike GFP, TAMRA exhibits $R$-dependent slow translational diffusion while maintaining $\alpha \sim 1$. A possible cause of the $R$-dependent slow diffusion is the increase in the tracer radius, $r$, due to the clustering of TAMRA molecules induced by the cell-sized confinement. To verify this hypothesis, we analyzed the rotational diffusion coefficient, $D_R$, which is expected to significantly decrease with an increase in $r$ as $D_R \sim 1/\eta r^3$ based on the Stokes–Einstein–Debye equation.[30] For this experiment, we need to separate the contributions of the translational and rotational diffusions, such that the translational decay time, $\tau_D$, should be considerably longer than the rotational decay time, $\tau_R$. We used the tracer molecule, R6G, for this rotational diffusion analysis since R6G has a similar size as TAMRA ($2r_{TAMRA} = 1.7$ nm $\sim 2r_{R6G} = 1.5$ nm), and reference values have been reported [31] for R6G but not for TAMRA.

Prior to the $D_R$ analysis inside dextran droplets, we confirmed that the effect of polarization of the excitation light is negligible in the diffusion measurements from the almost identical translational correlation time, $\tau_D$, in the bulk with or without the use of polarizers (Table S1 and section S1 in the SI). In addition, the values of $\tau_L$ and $\tau_R$ in the bulk without dextran are almost identical to the reference values.[32] Accordingly, we obtained $\tau_D$ inside the dextran droplets and plotted the normalized diffusion coefficients, $D/D_{bulk}$, against $R$ for systems with and without polarizers (Figure 2a). We again observe a slow translational diffusion of R6G inside small dextran droplets with $R < 20$ μm, as in the case of TAMRA (Figure 1e).

Accordingly, we evaluate the $D_R$ inside the dextran droplets based on the bulk values (Table S1 in the SI). The resulting $R$ dependence of $\tau_R$ and normalized $D_R/D_{R,\,bulk}$ are shown in Figure 2b. Even for small dextran droplets with $R < 20$ μm, which exhibit strongly slowed translational diffusion, $D_R/D_R$ is independent of $R$, maintaining a value almost equal to that in larger droplets and the bulk. This result suggests that the size of the tracer molecule, $r$ is essentially constant even for small droplets containing a highly concentrated dextran solution. Subsequently, we discuss why $D_R$ appears to have increased for small droplets with $R = 10$ μm.

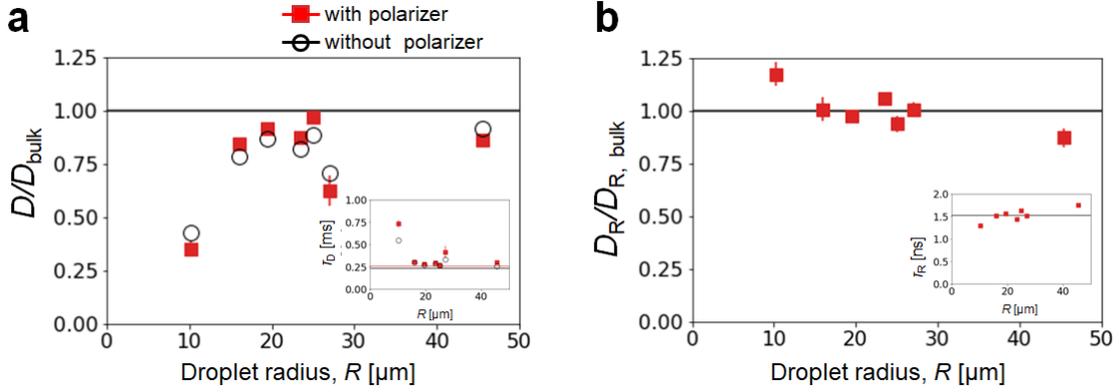

**Figure 2.** Droplet radius $R$ dependence of (a) normalized translational diffusion coefficient, $D/D_{bulk}$, and (b) normalized rotational diffusion coefficient, $D_R/D_{R,\,bulk}$, of R6G inside 400 mg/mL dextran droplets using excitation light with polarizers (closed red squares) and without polarizers (open black circles). The insets show the $R$ dependence of the (a) translational decay time, $\tau_D$, and (b) rotational decay time, $\tau_R$. The horizontal line shows their corresponding bulk values: (a) translational time analyzed with or without polarizers, $\tau_D = 0.26 \pm 0.01$ ms (ave. ± standard error (S.E.), N = 3) and $0.234 \pm 0.005$ ms (ave. ± S.E., N = 6) and (b) rotational time analyzed with polarizers, $\tau_R = 0.90 \pm 0.01$ ns (ave. ± S.E., N = 3).

**Chain-length polydispersity enhances slow diffusion**

Next, we consider if the heterogeneous distribution of dextran chains inside the cell-sized droplets was the cause of the $R$-dependent slow translational diffusion of TAMRA. We expect that the higher $M_w$ dextran chains are near the center of the droplet because of their lower wettability, ultimately inhibiting molecular transport in that region because of their higher viscosity and smaller mesh size, $\xi$.

Based on this hypothesis, we added shorter Dex10k polymers to the Dex150k solutions, expecting an enhanced localization of longer dextran chains at the droplet center, which should significantly decrease the $D$ of TAMRA inside the small droplets, as illustrated in Figure 3a. Although the fluorescence labeling of Dex10k and Dex150k indicates a uniform distribution of dextran chain lengths inside the droplets (Figure S2), the polymer chain lengths may still be inhomogeneously distributed at scales smaller than the resolution of fluorescence microscopy. To confirm that the shorter Dex10k have a higher membrane wettability than Dex150k, the interfacial tension at the droplet surface, $\gamma$, was measured using the pendant drop method (section S1 in the SI). Owing to the difficulty in analyzing the $\gamma$ for droplets containing a highly viscous solution of 400 mg/mL Dex150k, we analyzed the relaxation process for dextran droplets containing 53 mg/mL Dex150k and investigated the effects of adding shorter Dex10k chains at concentrations $C_{Dex10k}$ = 1.1 and 11 mg/mL (Figure S3). As shown in Figure 3b, an increase in $C_{Dex10k}$ decreased $\gamma$, showing that the shorter Dex10k indeed have a higher membrane wettability than Dex150k, as expected.

To establish a reference, we first analyzed the anomalous exponent, $\alpha$, and the diffusion coefficient, $D$, of the bulk solutions of a 400 mg/mL Dex150k solution with small amounts of Dex10k (0.1, 0.4, and 1.2 mg/mL) (< 0.5 wt% of total dextran). As shown in Figure 3a (i, iv), these small additions of short Dex10k did not alter the $\alpha$ or $D$ of the bulk solutions. We also confirmed that the solution viscosity was maintained by the addition of 1.2 mg/mL Dex10k (Table S2). Next, the dextran solutions were confined in droplets, and we measured the $\alpha$ and $D/D_{bulk}$ against the droplet radius, $R$, where the $D_{bulk}$ is the diffusion coefficient of TAMRA in a bulk solution of 400 mg/mL Dex150k without adding shorter Dex10k (the same data shown in Figure 1e). The $R$-dependence of $\alpha$ and $D/D_{bulk}$ is shown in Figure S4. To visualize the effect of the addition of Dex10k, we combined the data of $R \geq 20$ μm and $R < 20$ μm and plotted them against the concentration of Dex10k, $C_{Dex10k}$ (Figure 3c). For $\alpha$, the trend of the droplets is similar to that of the bulk, *i.e.*, $\alpha$ is almost unity independent of $R$ and $C_{Dex10k}$ (Figure 3c, ii–iii). In contrast, $D/D_{bulk}$ is less than unity for droplets with any $R$ (Figure 3c, v–vi). The decay of $D/D_{bulk}$ with increasing $C_{Dex10k}$ becomes particularly pronounced for small droplets with $R < 20$ μm, where it reaches less than half of the bulk value ($D/D_{bulk} < 0.5$).

These results strongly support the hypothesis that the slow translational diffusion of TAMRA ($2r/\xi \approx 0.7$–0.9) inside dextran droplets with $R < 20$ μm originates from the $M_w$-dependent (or length-dependent) membrane wetting of dextran polymers. Short and long polymers were preferentially distributed on the droplet surface and at the center, respectively.

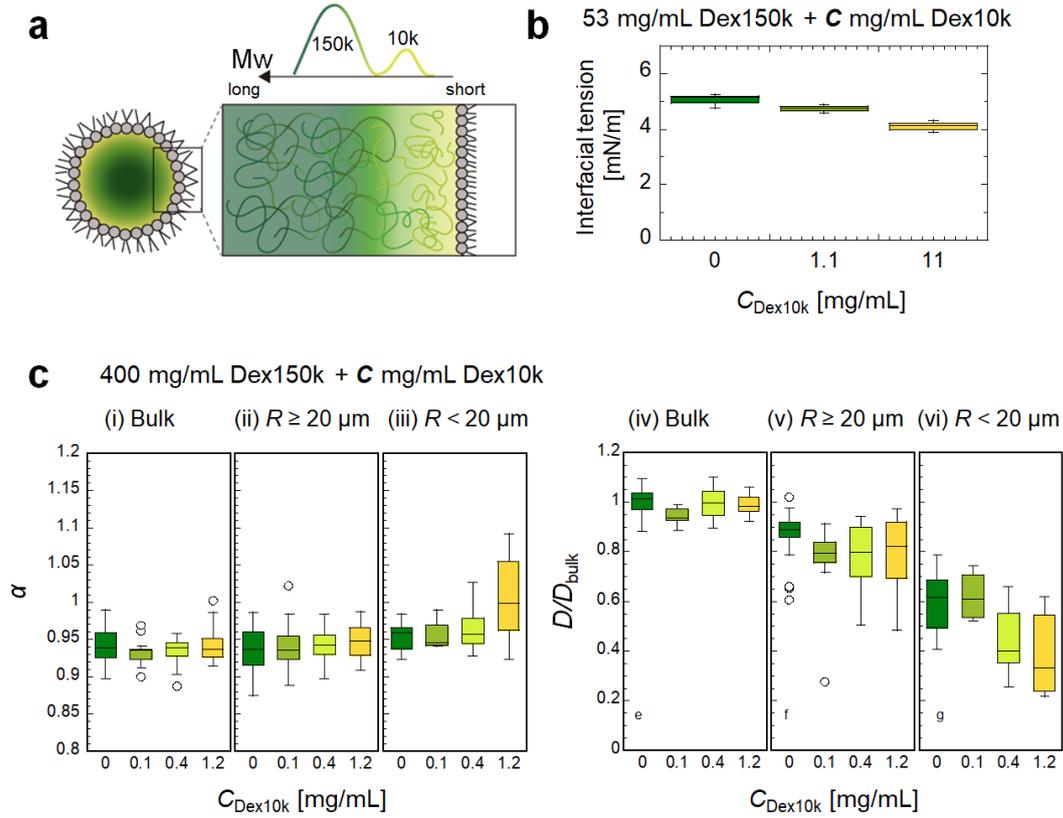

**Figure 3.** (a) Illustration of confinement-induced heterogeneous length distribution of dextran polymers. (b) Interfacial tension at the surface of droplets containing 53 mg/mL Dex150k without/with the addition of 1.1 or 11 mg/mL short Dex10k. (c) Anomalous exponent $\alpha$ (i-iii) and normalized diffusion coefficient $D/D_{bulk}$ of TAMRA (iv–vi) of the bulk (i, iv); of large droplets with $R \geq 20$ μm (ii, v); and small droplets with $R < 20$ μm (iii, vi) of 400 mg/mL Dex150k containing different concentrations of Dex10k (from left to right; 0, 0.1, 0.4, and 1.2 mg/mL). $D_{bulk}$ represents the $D$ of TAMRA in a bulk solution of 400 mg/mL Dex150k without Dex10k (the same data shown in Figure 1e). The error bars are S.D. The numbers of trials (N) corresponding to the added dextran concentrations (0, 0.1, 0.4, and 1.2 mg/mL) were 25, 13, 21, and 15 (for the bulk); 27, 18, 18, and 20 (for $R > 20$ μm); and 13, 4, 11, and 8 (for $R < 20$ μm).

**Heterogeneous distribution by chain length inside droplets**

To study the distribution and conformation of the confined polymers with a molecular-level resolution, we performed dissipative particle dynamics (DPD) simulations of Dex150k solutions confined in spherical droplets. Here, we used the same model as that in our recent computational study on PEG-dextran mixtures,[15] and we refer the reader to Ref.[15] and section S3 in the SI for technical details. In this model, each Dex150k chain is represented by a linear chain of 28 spherical beads connected by harmonic springs (an example of a polymer chain comprising 14 beads is shown in Figure 4a). To mimic the good solubility of dextran in water, we set the Flory–Huggins interaction parameter to $\chi = 0$. We investigate solutions at four different conditions of dispersity, i.e., the polydispersity index (PDI, the ratio between $M_w$ and number average molecular weight ($M_n$)) = 1.0, 1.3, 1.6, and 2.0 while fixing the monomer concentration of dextran at $C$ = 30 wt% (~400 mg/mL). The droplet was modeled as a spherical container with radius $R$ = 255 nm (Figure 4b). Although this size is approximately 40–200 times smaller than typical droplet sizes in experiments, the simulated droplets are large enough to capture the transition from the confinement-induced structuring near the droplet surface to the homogeneous distribution at the droplet interior, as demonstrated below.

As shown in Figure 4c, we plot the probability distribution, $P_{mono}(l)$, of finding a monomer at distance $l$ from the droplet center for monodisperse solutions (PDI = 1.0) and solutions with different PDI values, i.e., (i) PDI = 1.3 and (ii) PDI = 2.0 (data for PDI = 1.6 are shown in Figure S5), itemized by molecular weight. The distribution of the monodisperse case (PDI = 1.0) is shown as a red dotted line in each graph. These curves have been determined by averaging over 2000 independent simulation snapshots. In all cases, we can identify a distinct layering of monomers near the droplet surface, which decays near $l \approx 235$ nm in the monodisperse case; the width of this layered region is approximately 25 nm, which is consistent with the characteristic diameter of the Dex150k chains in our simulations, i.e., $2R_g \approx 26$ nm. For polydisperse solutions, this zone becomes narrower/broader as the molecular weight decreases/increases relative to 150k, respectively. Comparing the curves of the different $M_n$ windows, we can clearly see that the monomers from the shorter chains preferentially wet the droplet surface, which is consistent with the interfacial tension data shown in Figure 3b. Owing to this preferential surface wetting of short dextran chains, there is a slight surplus of long dextran chains near the droplet center. Furthermore, if we compare the solutions with different PDI values, we discover that the differences in the $P(l)$ of the different $M_n$ groups become slightly more pronounced with an increase in the PDI. After conducting the same analysis for the probability distribution of the polymer centers of masses, $P_{poly}(l)$, the same trends were observed (Figure 4d).

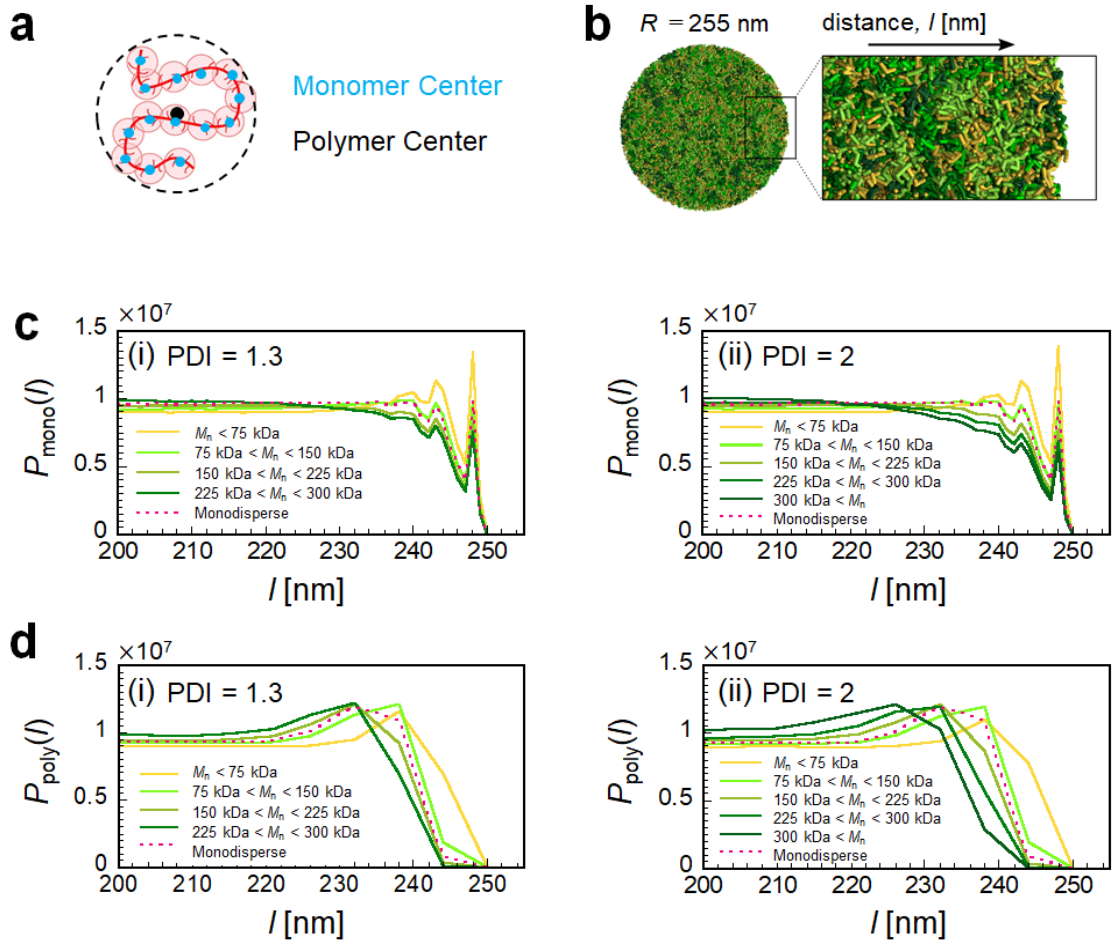

**Figure 4**: (a) Schematic representations of the model for a dextran chain (e.g., a chain of 14 beads). (b) Cross-sectional simulation snapshot of a dextran droplet with PDI = 2.0; chains are colored according to their molecular weight as in (c,d). (c, d) Probability distribution of finding a (c) monomer $P_{mono}(l)$ or (d) polymer $P_{poly}(l)$ at distance $l$ from the droplet center for polydisperse solutions with PDI = 1.3 and 2.0 (data for PDI = 1.6 is shown in Figure S5). The red dashed line shows the data of a monodisperse solution (PDI = 1.0).

We demonstrated that the macromolecular crowding inside W/O droplets with a radius $R < 20$ μm caused a significant slowing down of the translational diffusion, $D$, of small tracer molecules (TAMRA, $2r/\xi \approx 0.7$–$0.9$) compared with that of large droplets and in the bulk while the normalized rotational diffusion and anomalous exponent $\alpha$ remains almost unity. We rationalized this slow translational diffusion by the preferential wetting of short polymers at the droplet surface, which causes a surplus of long dextran chains at the droplet center (Figure 5). The most likely cause of the slow translational diffusion is the higher viscosity of longer polymers and their confinement-induced slow relaxation.

Notably, a slight increase was observed in the rotational diffusion of TAMRA as the droplet radius decreased (Figure 2b). In addition, $\alpha$ appeared to approach unity with an increasing concentration of short Dex10k (Figure 3c, iii). These changes could be interpreted as a slight increase in $\xi$; however, this change in $\xi$ was still sufficient to satisfy $2r/\xi \approx 1$, as observed for the size-dependent translational diffusion.

Although this $R$-dependent slow translational diffusion was not apparent for larger GFP tracer molecules ($2r/\xi \approx 2.3$–$2.9$), both $\alpha$ and $D$ slightly fell with a decrease in $R$. This trend can be explained by the strong trapping of large GFP molecules by the confined polymer chains, where polymer relaxation was also slowed by cell-size confinement.[25] This $R$-independence of large molecules is similar to the behavior of chromatin with a few micrometers in size as it diffuses through nuclei, following dynamic scaling regardless of the nuclear space size.[33] Therefore, the cell-sized confinement is likely a hidden parameter, which regulates molecular diffusion at intermediate length scales ($2r \approx \xi$), as shown in Figure 5.

The ability to control the spatial distribution of various polymers using differences in their surface wettability can be applied in various systems. For binary polymer blends of Dex70k and non-gelling gelatin, the interfacial tension decreases with the addition of short dextran [34], suggesting that the added short polymers are localized at the phase boundary to minimize the interfacial energy. Similarly, living cells have various surfaces, such as the cellular membrane, membrane organelle, and the surfaces of biomolecular condensates. Corresponding to this, the lengths of RNA and intrinsically disordered regions of RNA-binding proteins have been reported to be crucial for the emergence of LLPS and the physicochemical properties of the condensates.[35-37] Therefore, our findings on the wetting-induced heterogeneity of polymer chain length may contribute to the understanding of the heterogeneous distribution of molecules inside living cells.

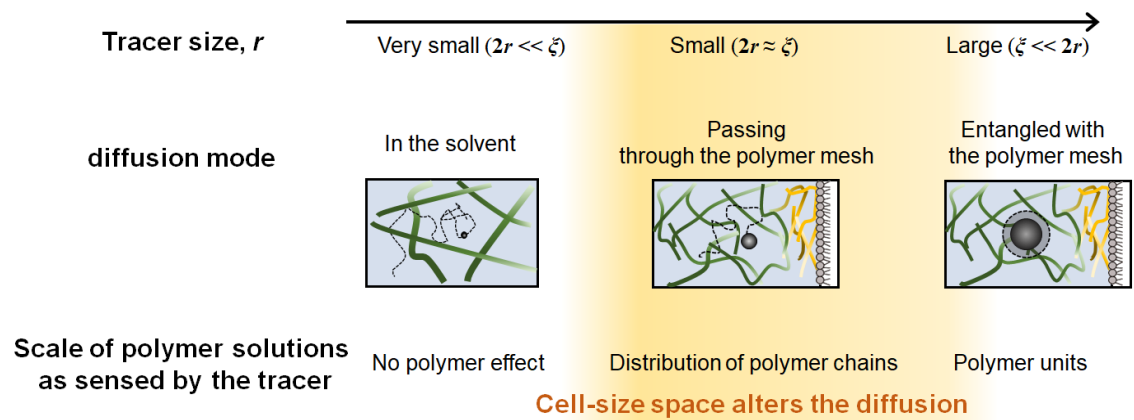

**Figure 5.** Possible mechanism of cell-sized space altering small-molecule diffusion through the polymer chains.

## EXPERIMENTAL SECTION
Experimental details are provided in section S1 in the Supporting Information

## ASSOCIATED CONTENT
**Supporting Information**
The Supporting Information is available free of charge at Materials and methods (section S1), Molecular simulation (section S2), estimation of correlation length in the dextran solution (section S3), Decay time of R6G in bulk solutions (Table 1), Viscosity of dextran solutions (Table 2), Concentration dependence on the correlation length for dextran solution (Figure S1), Fluorescence images of a dextran droplet (Figure S2), relaxation curves of interfacial tension γ on the surface of dextran droplets (Figure S3), droplet size dependence of TAMRA diffusion inside polydisperse dextran droplets (Figure S4), Probability distribution of monomer or polymer against the distance from the droplet center (Figure S5).

**Notes**
The authors declare no competing financial interest.

**Acknowledgements**
This research was funded by the Japan Society for the Promotion of Science (JSPS) KAKENHI (grant numbers 21H05871 and 22H01188 (M.Y.), 20K14425 (C.W.)) Japan Science and Technology Agency (JST) Program FOREST (JPMJFR213Y (M.Y.)) and ACT-X (JPMJAX191L (CW)). A.N. acknowledges funding provided by the Deutsche Forschungsgemeinschaft (DFG, German Research Foundation) through project number 470113688.

# Supporting Information

# Cell-sized confinements alter molecular diffusion in concentrated polymer solutions due to length-dependent wetting of polymers

Yuki Kanakubo[1], Chiho Watanabe[2], Johtaro Yamamoto[3], Naoya Yanagisawa[1], Arash Nikoubashman[4,5], Miho Yanagisawa[1,6,7*]

[1] Komaba Institute for Science, Graduate School of Arts and Sciences, The University of Tokyo, Komaba 3-8-1, Meguro, Tokyo 153-8902, Japan
[2] School of Integrated Arts and Sciences, Graduate School of Integrated Sciences for Life, Hiroshima University, Kagamiyama 1-7-1, Higashi-Hiroshima 739-8521, Japan
[3] Biomedical Research Institute, National Institute of Advanced Industrial Science and Technology (AIST), Central 6, Higashi 1-1-1, Tsukuba, Ibaraki 305-8568, Japan
[4] Institute of Physics, Johannes Gutenberg University Mainz, Staudingerweg 7, 55128 Mainz, Germany
[5] Department of Mechanical Engineering, Keio University, Hiyoshi 3-14-1, Kohoku, Yokohama 223-8522, Japan
[6] Graduate School of Science, The University of Tokyo, Hongo 7-3-1, Bunkyo, Tokyo 113-0033, Japan
[7] Center for Complex Systems Biology, Universal Biology Institute, The University of Tokyo, Komaba 3-8-1, Meguro, Tokyo 153-8902, Japan

**AUTHOR INFORMATION**
*Corresponding Author:
myanagisawa@g.ecc.u-tokyo.ac.jp (M. Y.)

**Contents**

## S1. Materials and Methods
**Materials**

For diffusive tracer molecules, we used three types of fluorescence molecules: 5-carboxytetramethylrhodamine (TAMRA) (Sigma-Aldrich Japan, Tokyo, Japan), Rhodamine 6G (R6G) (Tokyo Chemical Industry Co. Ltd., Tokyo, Japan), and Super folder green fluorescent protein (GFP) provided by Professor Kei Fujiwara of Keio University. As crowder molecules, we used the branched polymer dextran with average molecular weights ($M_w$) of 150 kDa and 10 kDa (Sigma-Aldrich Japan, Tokyo, Japan), here referred to as Dex150k and Dex10k, respectively. To analyze the diffusion of crowder dextran molecules, we used Rhodamine-labeled dextran with nominal $M_w$ = 100 kDa (RB-Dex100k; Sigma-Aldrich Japan), which was added to be 150 nM. These dextran polymers were dissolved in a phosphate buffered saline (PBS) buffer (137 mM NaCl, 8.1 mM $Na_2HPO_4$, 2.68 mM KCl, 1.47 mM $KH_2PO_4$, pH 7.4 ± 0.2) using 10x PBS buffer (314-90185, Nippon Gene Co., Ltd., Toyama, Japan). To visualize the length distribution of dextran chains, 0.1 or 1.2 mg/mL Rhodamine-labeled dextran with nominal $M_w$ of 10 kDa (RB-Dex10k) and 0.4 mg/mL Fluorescein Isothiocyanate labeled dextran with nominal $M_w$ of 150kDa (FITC-Dex150k; Sigma-Aldrich Japan) were added to 400 mg/mL Dex150k samples. To prepare water-in-oil (W/O) droplets of the dextran solutions, 1,2-Dioleoyl-sn-glycero-3-phosphocholine or 1-palmitoyl-2-oleoyl-sn-glycero-3-phosphocholine (PC; NOF Corporation, Tokyo, Japan) and mineral oil (kinetic viscosity of 14−17 $mm^2s^{−1}$ at 38 °C) from Nacalai Tesque (Kyoto, Japan) were used as surfactant and oil phase, respectively. Mineral oils were treated with molecular sieves (4A; AS ONE, Osaka, Japan); all other materials were used without purification.

**Preparation of polymer droplets**

Following our previous reports about the slow diffusion inside Dex150k droplets, we fixed the concentration of Dex150k to be 400 mg/mL[1]. Samples with 400 mg/mL Dex150k in PBS solution containing 20-50 nM fluorescent molecules (i.e., TAMRA or GFP) were prepared. 1-2 mM PC lipids in mineral oil were prepared by mixing lipids-in-chloroform with mineral oil and evaporating the chloroform at 70°C overnight. Then, we added the polymer solution to the lipid-oil solution at a volume ratio of 1.5:100 (w/o). By emulsification via pipetting, we prepared droplets with radii $R$ ranging from 5−70 μm. The droplet solution was sandwiched between two cover glasses (No. 1, 0.12–0.17 mm, Matsunami Glass Co., Kishiwada, Osaka, Japan) with spacers of ~0.1 mm thickness (double sided sticky tape; Askul, Tokyo, Japan) or 25 μL flame (Gene Frame; Thermo Scientific™, Tokyo, Japan).

**Interfacial tension on the surface of dextran droplets**

The interfacial tension on the surface of PC droplets containing dextran solutions was analyzed at approximately 20 °C using the pendant drop method (DM-501, Kyowa Interface Science Co., Saitama, Japan). PC in a mineral oil solution of ~ 1 mg/mL was used as the oil phase. We note that it was difficult to analyze the interfacial tension of droplets containing 400 mg/ml dextran solution because of their very high viscosity.[1] Therefore, we analyzed the relaxation process for 53 mg/mL Dex150k, and the effect of shorter Dex10k addition, $C_{Dex10k}$ = 1.1, 11 mg/mL. There relaxation process was fitted with an exponential curve, $\exp(-\tau/t)$, where $\tau$ is the correlation time. Here, we compared the interfacial tension $\gamma$ for some different dextran compositions at $t = \tau$.

**Fluorescence observation of dextran droplets**

Fluorescence images of the bulk solutions and droplets were obtained using a confocal laser scanning microscope (IX83, FV1200; Olympus Inc., Tokyo, Japan) equipped with a water immersion objective lens (UPLSAPO 60XW, Olympus Inc.). FITC-Dex150k and RB-Dex10k were excited at wavelengths of 473 and 559 nm and detected in the ranges of 490–540 and 575–675 nm, respectively. The fluorescence intensity was analyzed using Fiji software (National Institute of Health (NIH), USA).[2]

**Viscosity measurement of dextran solution**

Static viscosity (viscosity [Pa·s] x density [kg/m3]) of 400 mg/mL Dex150k solution with or without 1.2 mg/mL Dex10k was measured by using a sinewave vibro viscometer (A&D, SV-10H, Japan). To obtain the viscosity, the value was divided by the solution density by using a density meter (DMA 1001, Anton Paar GmbH) with a resolution of 0.05 mg/mL.

**Translational diffusion analysis**

Translational diffusion was analyzed by using point fluorescence correlation spectroscopy (point FCS) (FV1200 confocal microscope integrated with a molecular diffusion package; Olympus Inc. Tokyo, Japan) equipped with 473 and 559 nm diode lasers. A water immersion objective lens (UPLSAPO 60XW, Olympus Inc.) was used to perform point FCS with minimal laser power (~0.3 µW), to minimize refractive index (RI) mismatch and temperature change during laser irradiation. Molecular diffusion inside the droplets was measured at the center of a droplet attached to the glass and at a height of ~5 µm from the glass surface to eliminate artifacts due to RI mismatch through the surrounding oil or concentrated dextran solution.[3] TAMRA was excited at a wavelength of 559 nm and detected in the range 575–675 nm. R6G and GFP were excited by the 473 nm laser and detected in the 490–590 nm range. We derived the autocorrelation functions via the accompanied software of the microscope. The illumination volumes were calibrated with the TAMRA and R6G solutions (< 100 nM) by using the same diffusion coefficient value of 280 µm²/s. The $xy$ radii of the illumination volume, $w_0$ and its ratio with respect to the z-direction height $w_z$, $s$ (= $w_z/w_0$, structure factor) were 0.20–0.22 μm and 4–5 for TAMRA and 0.17 μm and 6 for R6G, respectively. We analyzed the translational diffusion behaviors by fitting ACFs $G_D(\tau)$ as a function of the correlation time $\tau$, with a Brownian diffusion model or fractional Brownian motion (fBM) model with an anomalous exponent parameter, $\alpha$ [4-5]:

$$G_D(\tau) = \frac{1}{N}\left(1 + \left(\frac{\tau}{\tau_D}\right)^\alpha\right)^{-1}\left(1 + \frac{1}{s^2}\left(\frac{\tau}{\tau_D}\right)^\alpha\right)^{-\frac{1}{2}} \quad (1)$$

where $N$ is the number of fluorescent molecules, $\alpha$ is the anomalous exponent, $\tau_D$ is the characteristic decay time. As for Brownian diffusion mode, $\alpha$ is fixed to 1. From $\tau_D$ and $w_0$, the diffusion coefficient $D_\alpha$ can be derived as follows [4]:

$$D_\alpha = \frac{w_0^2}{4\tau_D^\alpha} \quad (2)$$

**Rotational diffusion analysis**

Rotational diffusion measurements of a fluorophore R6G at the center of a droplet were performed with polarization-dependent fluorescence lifetime correlation spectroscopy (FLCS) using TCS SP8 STED 3X (Leica Microsystems GmbH). By placing a polarizer in the optical path of the FLCS device, we measured both rotational and translational diffusion by simultaneously measuring time-resolved fluorescence anisotropy and FCS. Here we refer to this as pol-FLCS. A water immersion objective (HC PL APO 86x/1,20 W motCORR STED white) was used for the measurements with a low laser intensity where the triplet effect is negligible. R6G was excited with wavelength of 488 nm and detected in the range of 500-698 nm. The structure factor $s$ of the illumination volume was 8.2. Similar to the above-mentioned point FCS experiments, we analyzed the molecular diffusion at the center of adhered droplets on the glass and at a height of ~3 μm from the bottom glass surface. According to the pol-FLCS results, ACF $G_D(\tau)$ and fluorescence decay $I(\tau)$ were derived using the integrated software in time ranges of 5 μs to 0.5 s and 4 ps to 13 ns, respectively. Translational diffusion coefficient $D$ and rotational diffusion coefficient $D_R$ were calculated from their characteristic decay time, $\tau_D$ and $\tau_R$, and lifetime of the fluorescence molecule, $\tau_L$. Here we hypothesize the rotational diffusion is Brownian and then fitted $G_D(\tau)$ and $I(\tau)$ with the Brownian diffusion model because the time scale of $\tau_R$ is ns, which is three orders of magnitude shorter than $\tau_D$.

First, fluorescence lifetime $\tau_L$ measurement is performed using circularly polarized excitation light (without polarizer). The fluorescence decay $I(\tau)$ is fitted with an exponential decay: [6-7]

$$I(\tau) = k\ \exp\left(-\frac{\tau}{\tau_L}\right), \quad (3)$$

where $k$ is the amplitude of the exponential decay. Next, $\tau_L$ measurement is performed by using linearly polarized excitation light with a polarizer. $I(t)$ only from the parallel component to the polarized excitation light is fitted by [8]

$$I(\tau) = k\ \exp\left(-\frac{\tau}{\tau_L}\right)\left[1 + 2r_0\exp\left(-\frac{\tau}{\tau_R}\right)\right] \tag{4}$$

where $r_0$ is the initial anisotropy of a fluorophore. R6G is a phosphor with S1→S0 absorption. Therefore, the transition moment vectors are usually close to parallel, so we fixed $r_0 = 0.4$ (the maximum value) in this analysis.[9] By the experiments using with or without the polarizer, we can obtain the values of $\tau_L$ and $\tau_R$.

## S2. Estimation of correlation length $\xi$ in dextran solution

The correlation length $\xi$ in condensed dextran solution were calculated from following equations.[1, 10-12] The radius of gyration $R_g$ of dextran in dilute solution, as well as other liner polymers, is expressed a function of $M_w$. Using the reported experimental values as a reference, [10-11, 13] here we estimated $R_g$ of Dex150k and Dex10k as 10-12 nm and 2.6-3.2 nm, respectively.

Using the $R_g$, the overlap concentration $c^*$ can be derived as follows,[10-11]

$$c* = \frac{M_w}{N_A}\left(\frac{4}{3}\pi R_g^3\right)^{-1} \tag{5}$$

where $N_A$ is Avogadro's constant. For Dex150k, $c^*$ is approximately 34~60 mg/mL. With $R_g$ and $c^*$, we can estimate $\xi$ at a polymer concentration $c$ ($> c^*$) based on the experimentally reported relationship,[12]

$$\xi = R_g\left(\frac{c}{c*}\right)^{-\frac{3}{4}} \quad \text{nm}. \tag{6}$$

As an example of Dex150k with $R_g = 10$ nm, $c^* = 55$ mg/mL, the $c$ dependence of $\xi$ is shown in Figure S1. In the experiments, 400 mg/mL was adopted, whose mesh size is estimated to be 1.9−2.4 nm, which is comparable to the diameters of three different tracer molecules used, *i.e.,* TAMRA ($2r = 1.7$ nm), R6G ($2r = 1.5$ nm), and GFP ($2r = 5.6$ nm).

## S3. Molecular simulation

We employed dissipative particle dynamics (DPD) simulations of a bead-spring model, where each Dex150k polymer was represented as a linear chain consisting of 28 spherical beads of diameter $a \approx 6.0$ nm, following the mapping established in our previous work.[14] Water molecules are grouped as well into spherical beads of diameter $a$, thus representing about 3700 water molecules per bead. All DPD particles interact through soft, purely repulsive forces

$$\mathbf{f}_c(r) = \begin{cases} A_{ij}(1 - r/a)\hat{\mathbf{r}}, & r < a \\ 0, & r \geq a \end{cases} \quad (7)$$

where $\hat{\mathbf{r}}$ is the unit vector connecting particles $i$ and $j$, and $r$ is the distance between them. The parameter $A_{ij}$ controls the strength of the repulsion, which is set to $A_{ij} = 25 k_B T/a$ for all interaction pairs, resulting in good solvent conditions for the dispersed polymers. Beads in a dextran chain are connected via harmonic springs with force

$$\mathbf{f}_b(r) = -k\mathbf{r} \quad (8)$$

with spring constant $k = 4k_B T/a^2$.

In addition to the conservative forces, all particles are subjected to pairwise dissipative and random forces

$$\mathbf{f}_d(r) = -\gamma_{ij}\omega(r)(\hat{\mathbf{r}} \cdot \Delta\mathbf{v})\hat{\mathbf{r}} \quad (7)$$

$$\mathbf{f}_r(r) = [\gamma_{ij}\omega(r)]^{\frac{1}{2}} \zeta \hat{\mathbf{r}} \quad (9)$$

where $\gamma_{ij} = 4.5(mk_B T)^{1/2} a$ is the drag coefficient, $\Delta\mathbf{v}$ is the velocity difference between the two particles, and $\zeta$ is a uniformly distributed random number drawn for each particle pair, with $\langle \zeta(t) \rangle = 0$ and $\langle \zeta(t)\zeta(t') \rangle = 2k_B T \delta(t-t')$ to fulfill the fluctuation-dissipation theorem. We use the standard DPD weight function

$$\omega(r) = \begin{cases} \left(1 - \dfrac{r}{a}\right)^2, & r < 0 \\ 0, & r \geq 0 \end{cases} \quad (10)$$

To mimic droplet confinement, we placed the DPD particles in a spherical container with radius $R = 42.8a$, which interacts with the DPD particles through a purely repulsive Weeks-Chandler-Andersen (WCA) potential[15]

$$U_{\text{WCA}}(l) = \begin{cases} -4k_B T\left[\left(\dfrac{a}{R-l}\right)^{12} - \left(\dfrac{a}{R-l}\right)^6 + \dfrac{1}{4}\right], & R - l < 2^{\frac{1}{6}}a \\ 0, & R - l \geq 2^{\frac{1}{6}}a \end{cases} \quad (11)$$

where $l$ is the distance between the center of a bead and the droplet center. The total particle number density was set to $\rho = 3a^{-3}$, and the equations of motion were integrated using a time step of $\Delta t = 0.02\tau$, with simulation time unit $\tau$. All simulations were performed using the HOOMD-blue simulation package (v. 2.9.7).[16] Each simulation run for at least $10^7$ timesteps, and we performed two independent simulations for each dispersity.

**Supplemental figures**

**Table S1. Decay time of R6G in bulk solutions analyzed by using pol-FLCS**

Decay times for translational diffusion $\tau_D$ and rotational diffusion $\tau_R$, and lifetime of R6G with $2r$ ~1.5 nm in a bulk solution with or without 400 mg/mL Dex150k. $\tau_D$ are analyzed with or without the polarizer.

|  |  | Control in PBS buffer (Ave. ± S.E.) | 400 mg/mL Dex150k (Ave. ± S.E.) |
|---|---|---|---|
| $\tau_D$ [ms] | without Polarizer | 0.028 ± 0.001 (N = 3) | 0.256 ± 0.010 (N = 3) |
|  | with Polarizer | 0.028 ± 0.001 (N = 3) | 0.234 ± 0.005 (N = 6) |
| $\tau_L$ [ms] | without Polarizer | 3.747 ± 0.003 (N = 3) | 3.699 ± 0.003 (N = 6) |
| $\tau_R$ [ns] | with Polarizer | 1.005 ± 0.004 (N = 3) | 1.520 ± 0.007 (N = 3) |

**Table S2. Viscosity of Dex150k with or without Dex10k**

Viscosity of a solution of 400 mg/mL Dex150k with or without 1.2mg/mL Dex10k at 20.3 ± 0.4 °C.

|  | 400 mg/mL Dex150k (Ave. ± S.E.) | 400 mg/mL Dex150k + 1.2 mg/mL Dex10k (Ave. ± S.E.) |
|---|---|---|
| Viscosity [mPa sec] | 869 ± 4 (N = 5) | 820 ± 4 (N = 3) |

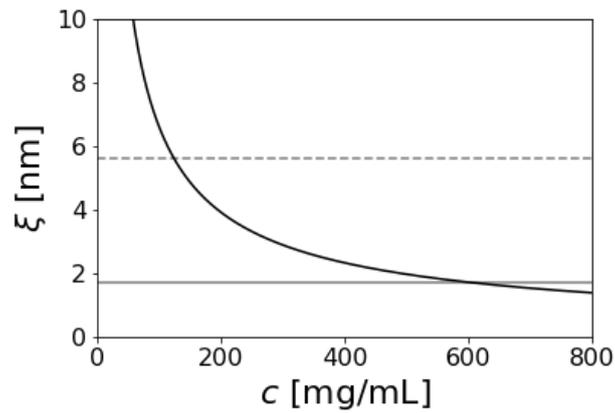

**Figure S1.** Concentration dependence on the correlation length $\xi$ for Dex150k solution in the case of $R_g$ = 10 nm, $c^*$ = 55 mg/mL. The horizontal lines indicate the diameter of TAMRA (solid line; $2r$ = 1.7 nm), and GFP (dotted line; $2r$ = 5.6 nm), respectively.

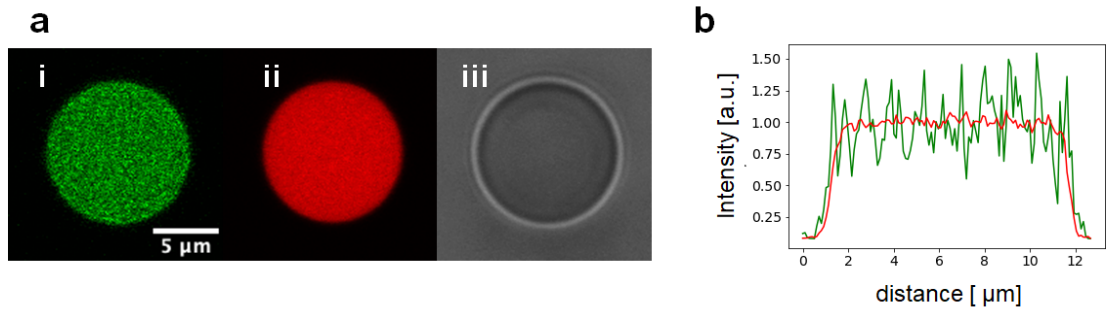

**Figure S2.** (a) (i, ii) Fluorescence images of a PC droplet with $R$ = 5.5 μm containing 400 mg/mL Dex150k, 0.4 mg/mL FITC-Dex150k (i, green) and 1.2 mg/mL RB-Dex10k (ii, red) and (iii) the transmission image, respectively. (b) Fluorescence intensity profile of RB-Dex10k (red) and FITC-Dex150k (green) along equatorial plane of the droplet. The intensities are normalized by the average value at the droplet center (distance = 2~11 μm).

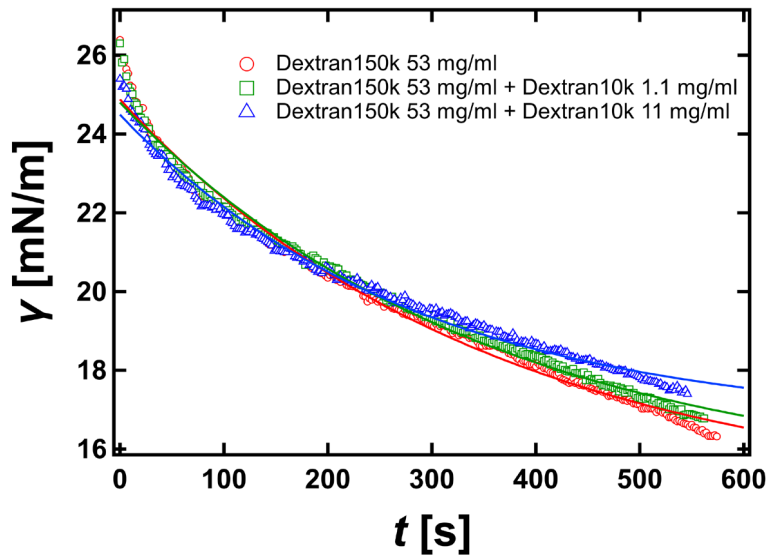

**Figure S3.** Relaxation curves of interfacial tension γ on the surface of PC droplets containing 53 mg/mL Dex150k solution only (red circle), and with 1.1 mg/ml Dex10k (green square) or 11 mg/ml Dex10k (blue triangle), respectively. Each solid line indicates exponential fitting by exp(-τ/t), where τ is the correlation time. Here, we obtained the value of γ at $t = τ$.

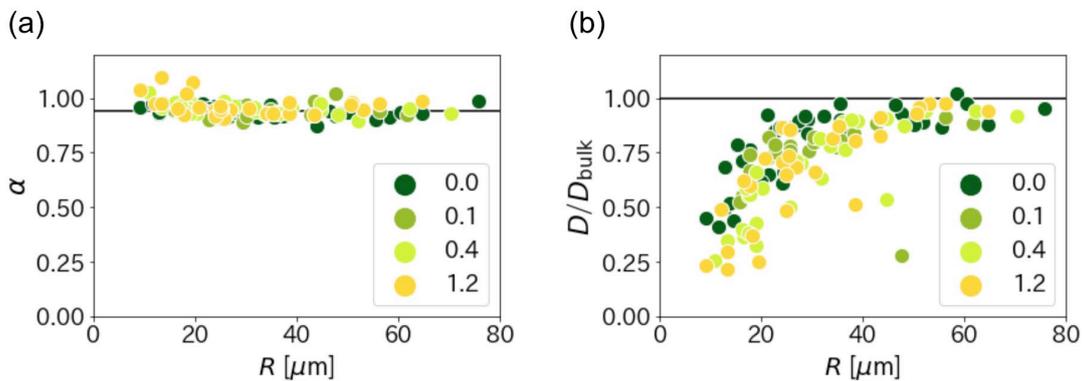

**Figure S4.** Effect of molecular weight polydispersity of dextran solution on TAMRA diffusion inside dextran droplets with a radius, $R$. Anomalous diffusion exponent $α$ (a) and normalized diffusion coefficient $D/D_{bulk}$ (b) for 400 mg/mL Dex150k with different concentration of Dex10k. The legend shows the added amount of Dex10k in unit of mg/mL. The bulk values for 400 mg/mL Dex150k only is (a) $α_{bulk} = 0.94 ± 0.02$ (Ave. ± S. D., N = 25) shown as a black solid line and (b) $D_{bulk} = 38.2 ± 2.1$ μm$^2$/sec (Ave. ± S. D., N = 25).

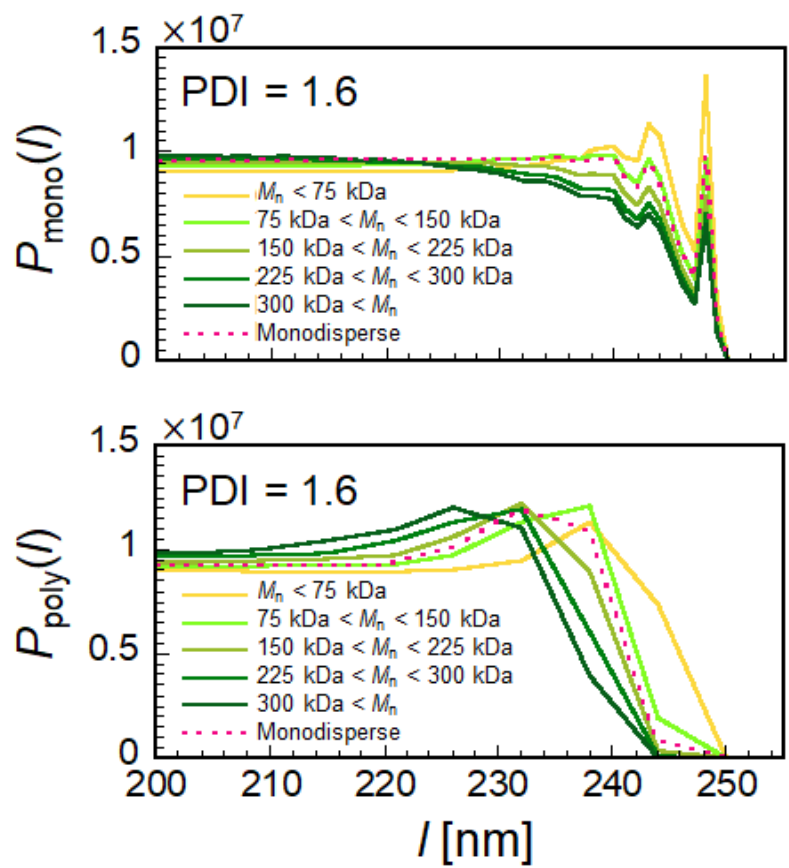

**Figure S5**. Probability distribution $P(l)$ of finding a monomer (top) and polymer (bottom) center of mass at a distance $l$ from the droplet center for polydisperse solutions with PDI = 1.6 (data for PDI=1.3, 2.0 are shown in Figures 4c, 4d).